\PassOptionsToPackage{unicode}{hyperref}
\PassOptionsToPackage{hyphens}{url}
\PassOptionsToPackage{dvipsnames,svgnames,x11names}{xcolor}
\documentclass[
  12pt]{article}

\usepackage{amsmath,amssymb}
\usepackage{iftex}
\ifPDFTeX
  \usepackage[T1]{fontenc}
  \usepackage[utf8]{inputenc}
  \usepackage{textcomp} % provide euro and other symbols
\else % if luatex or xetex
  \usepackage{unicode-math}
  \defaultfontfeatures{Scale=MatchLowercase}
  \defaultfontfeatures[\rmfamily]{Ligatures=TeX,Scale=1}
\fi
\usepackage{lmodern}
\ifPDFTeX\else  
\fi
\IfFileExists{upquote.sty}{\usepackage{upquote}}{}
\IfFileExists{microtype.sty}{% use microtype if available
  \usepackage[]{microtype}
  \UseMicrotypeSet[protrusion]{basicmath} % disable protrusion for tt fonts
}{}
\makeatletter
\@ifundefined{KOMAClassName}{% if non-KOMA class
  \IfFileExists{parskip.sty}{%
    \usepackage{parskip}
  }{% else
    \setlength{\parindent}{0pt}
    \setlength{\parskip}{6pt plus 2pt minus 1pt}}
}{% if KOMA class
  \KOMAoptions{parskip=half}}
\makeatother
\usepackage{xcolor}
\makeatletter
\ifx\paragraph\undefined\else
  \let\oldparagraph\paragraph
  \renewcommand{\paragraph}{
    \@ifstar
      \xxxParagraphStar
      \xxxParagraphNoStar
  }
  \newcommand{\xxxParagraphStar}[1]{\oldparagraph*{#1}\mbox{}}
  \newcommand{\xxxParagraphNoStar}[1]{\oldparagraph{#1}\mbox{}}
\fi
\ifx\subparagraph\undefined\else
  \let\oldsubparagraph\subparagraph
  \renewcommand{\subparagraph}{
    \@ifstar
      \xxxSubParagraphStar
      \xxxSubParagraphNoStar
  }
  \newcommand{\xxxSubParagraphStar}[1]{\oldsubparagraph*{#1}\mbox{}}
  \newcommand{\xxxSubParagraphNoStar}[1]{\oldsubparagraph{#1}\mbox{}}
\fi
\makeatother

\usepackage{longtable,booktabs,array}
\usepackage{calc} % for calculating minipage widths
\usepackage{etoolbox}
\makeatletter
\patchcmd\longtable{\par}{\if@noskipsec\mbox{}\fi\par}{}{}
\makeatother
\IfFileExists{footnotehyper.sty}{\usepackage{footnotehyper}}{\usepackage{footnote}}
\makesavenoteenv{longtable}
\usepackage{graphicx}
\makeatletter
\def\maxwidth{\ifdim\Gin@nat@width>\linewidth\linewidth\else\Gin@nat@width\fi}
\def\maxheight{\ifdim\Gin@nat@height>\textheight\textheight\else\Gin@nat@height\fi}
\makeatother
\setkeys{Gin}{width=\maxwidth,height=\maxheight,keepaspectratio}
\makeatletter
\def\fps@figure{htbp}
\makeatother

\makeatletter
\@ifpackageloaded{caption}{}{\usepackage{caption}}
\AtBeginDocument{%
\ifdefined\contentsname
  \renewcommand*\contentsname{Table of contents}
\else
  \newcommand\contentsname{Table of contents}
\fi
\ifdefined\listfigurename
  \renewcommand*\listfigurename{List of Figures}
\else
  \newcommand\listfigurename{List of Figures}
\fi
\ifdefined\listtablename
  \renewcommand*\listtablename{List of Tables}
\else
  \newcommand\listtablename{List of Tables}
\fi
\ifdefined\figurename
  \renewcommand*\figurename{Figure}
\else
  \newcommand\figurename{Figure}
\fi
\ifdefined\tablename
  \renewcommand*\tablename{Table}
\else
  \newcommand\tablename{Table}
\fi
}
\@ifpackageloaded{float}{}{\usepackage{float}}
\floatstyle{ruled}
\@ifundefined{c@chapter}{\newfloat{codelisting}{h}{lop}}{\newfloat{codelisting}{h}{lop}[chapter]}
\floatname{codelisting}{Listing}

\makeatother
\makeatletter
\@ifpackageloaded{caption}{}{\usepackage{caption}}
\@ifpackageloaded{subcaption}{}{\usepackage{subcaption}}
\makeatother

\ifLuaTeX
  \usepackage{selnolig}  % disable illegal ligatures
\fi
\usepackage[]{natbib}
\usepackage{bookmark}

\IfFileExists{xurl.sty}{\usepackage{xurl}}{} % add URL line breaks if available
\hypersetup{
  pdftitle={Going With the Flow: Normalizing Flows for Gaussian Process Regression under Hierarchical Shrinkage Priors},
  pdfauthor={Peter Knaus},
  pdfkeywords={Variational inference, Variable selection, Covariance shrinkage, Triple gamma shrinkage prior},
  colorlinks=true,
  linkcolor={blue},
  filecolor={Maroon},
  citecolor={Blue},
  urlcolor={Blue},
  pdfcreator={LaTeX} 
}
\newcommand{\anon}{1}

\newcommand{\Gamfun}[1]{\Gamma (#1)}
\newcommand{\Uhyp}[1]{U \left(#1\right)}
\newcommand{\Betafun}[1]{B (#1)}
\usepackage{algorithm}
\usepackage{algpseudocode}

\begin{document}

\def\spacingset#1{\renewcommand{\baselinestretch}%
{#1}\small\normalsize} \spacingset{1}

%%%%%%%%%%%%%%%%%%%%%%%%%%%%%%%%%%%%%%%%%%%%%%%%%%%%%%%%%%%%%%%%%%%%%%%%%%%%%%

\if1\anon
{
  \title{\bf Going With the Flow: Normalizing Flows for Gaussian Process Regression under Hierarchical Shrinkage Priors}
  \author{Peter Knaus\thanks{This research was funded in whole or in part by the Austrian Science Fund (FWF) 10.55776/J4806.}\hspace{.2cm}\\
    Department of Statistics, Harvard University\\
    Institute for Statistics and Mathematics,  WU Vienna}
  \maketitle
} \fi

\if0\anon
{
  \bigskip
  \bigskip
  \bigskip
  \begin{center}
    {\LARGE\bf Going With the Flow: Normalizing Flows for Gaussian Process Regression under Hierarchical Shrinkage Priors}
\end{center}
  \medskip
} \fi

\bigskip
\begin{abstract}
  Gaussian Process Regression (GPR) is a powerful tool for nonparametric regression, but its application in a fully Bayesian fashion in high-dimensional settings is hindered by two primary challenges: the difficulty of variable selection and the computational burden, which is particularly acute in fully Bayesian inference. This paper introduces a novel methodology that combines hierarchical global-local shrinkage priors with normalizing flows to address these challenges. The hierarchical triple gamma prior offers a principled framework for inducing sparsity in high-dimensional GPR, effectively excluding irrelevant covariates while preserving interpretability and flexibility. Normalizing flows are employed within a variational inference framework to approximate the posterior distribution of parameters, capturing complex dependencies while ensuring computational scalability. Simulation studies demonstrate the efficacy of the proposed approach, outperforming traditional maximum likelihood estimation and mean-field variational methods, particularly in high-sparsity and high-dimensional settings. This is also borne out in an application to binding affinity ($\text{pIC}_{50}$) measurements for small molecules targeting $\beta$-secretase-1 (BACE-1). The results highlight the robustness and flexibility of hierarchical shrinkage priors and the computational efficiency of normalizing flows for Bayesian GPR. This work provides a scalable and interpretable solution for high-dimensional nonparametric regression, with implications for sparse modeling and posterior approximation in broader Bayesian contexts.
\end{abstract}

\noindent%
\vfill

\newpage
\spacingset{1.8} % DON'T change the spacing!

\section{Introduction}

    Gaussian process regression (GPR) is a widely used tool in nonparametric regression, valued for its theoretical rigor. It achieves near-minimax rates of convergence to a true underlying function \citep{vanderVaart2009}, all while retaining a higher degree of interpretability than fully black box methods. However, two major challenges arise in practice. First, selecting the hyperparameters that govern the covariance structure becomes increasingly complex as the number of predictors grows, requiring variable selection to distinguish influential covariates from noise. Second, estimating GPR models, particularly within a fully Bayesian framework, is computationally demanding. The cost of evaluating the marginal likelihood scales cubically with the number of observations, making standard approaches like Markov chain Monte Carlo (MCMC) prohibitively expensive for even moderately sized datasets.

    Several methods have been proposed for variable selection in GPR. The most well-known is automatic relevance determination (ARD) by \cite{neal1996bayesian}, which identifies covariate importance through maximum-likelihood estimates of the parameters but does not impose sparsity explicitly. Penalized approaches, such as those by \cite{yan2010sparse} and \cite{yi2011penalized}, address this by enforcing sparsity directly through, for example, the $\ell_1$ penalty. Alternatively, sensitivity-based methods \citep[e.g.,][]{piironen2016projection, paananen2019variable} evaluate the predictive changes when covariates are removed, while dimension-reduction techniques \citep[e.g.,][]{park2022variable, tripathy2016gaussian, liu2017dimension} focus on finding lower-dimensional representations of the covariates, often at the expense of interpretability.

    Despite being couched in Bayesian terminology, all previously mentioned methods for variable selection in GPR are not fully Bayesian, as they yield only point estimates for hyperparameters and lack the full uncertainty quantification inherent to Bayesian analysis. This limitation can result in overconfidence in predictions and reduced out-of-sample performance. Fully Bayesian approaches, while less common, aim to address this. For example, \cite{chen2010bayesian} and \cite{savitsky2011variable} use spike-and-slab priors \citep{mitchell1988bayesian, george1993variable, george1997approaches} to induce sparsity, providing posterior inclusion probabilities for each covariate. However, these methods become computationally expensive in higher dimensions. Others, such as \cite{vo2017sparse}, adopt computationally tractable alternatives like the horseshoe prior \citep{carvalho2010horseshoe} to induce sparsity in additive GPR frameworks. More recently, \cite{tang2023hierarchical} applied cumulative shrinkage priors \citep{legramanti2020bayesian} using the Karhunen-Lo\`eve decomposition \citep{alexanderian2015brief}, which embeds a hierarchy of effects but sacrifices interpretability.

    While theoretically more attractive, fully Bayesian approaches such as MCMC often particularly suffer from the high computational cost of evaluating the marginal likelihood. Consider single-move Metropolis-Hastings updates. Each parameter is updated sequentially, with each update requiring an evaluation of the marginal likelihood, making high-dimensional applications infeasible. Even more efficient samplers, like Hamiltonian Monte Carlo (HMC), can remain prohibitive due to the computational cost of evaluating gradients at each leapfrog step. Variational inference offers a potential solution by reducing computational demands, but many widely used implementations rely on mean-field approximations. This choice of variational family limits expressivity, making it virtually impossible to accurately capture complex posterior distributions. Taken together, these computational challenges constrain the applicability of fully Bayesian GPR to small datasets.

    This paper tackles the two aforementioned problems simultaneously. Recent insights from the hierarchical global-local shrinkage literature are introduced into the GPR framework, providing an interpretable and flexible approach to variable selection. Additionally, normalizing flows \citep{rezende2015variational} are employed within a variational inference framework to flexibly approximate the posterior distribution of the hyperparameters, capturing dependencies in the posterior while ensuring scalability and maintaining proper uncertainty quantification. Together, these advances enable a principled Bayesian approach to variable selection in GPR, yielding flexible, interpretable models that remain computationally tractable in higher dimensions. To facilitate practical application, all methods introduced in this paper are implemented in the R package \texttt{shrinkGPR} \citep{shrinkGPR}, which is available on CRAN.

    The remainder of this paper is structured as follows. Section~\ref{sec:shrinkage} introduces hierarchical global-local shrinkage priors for GPR. Section~\ref{sec:estim} provides an overview of normalizing flows and how they can be used to approximate the posterior distribution of the parameters. Section~\ref{sec:sim_study} presents a simulation study to assess the performance of the proposed method. Section~\ref{sec:app} applies the method to a dataset containing binding affinity ($\text{pIC}_{50}$) measurements for molecules targeting $\beta$-secretase-1 (BACE-1), a key enzyme implicated in Alzheimer's disease. Finally, Section~\ref{sec:conclusion} concludes.

\section{Hierarchical Shrinkage in Gaussian Process Regression} \label{sec:shrinkage}

This work considers the following multivariate nonparametric regression framework
\begin{equation}
    y_i = f(\mathbf{x}_i) + \epsilon_i, \quad \epsilon_i \overset{\hphantom{\text{iid}}}{\sim} \mathcal{N}(0, \sigma^2),
\end{equation}
where $f$ is an unknown mean regression function to be estimated from the data $\mathcal{D} = \{(\mathbf{x}_i, y_i)\}_{i=1}^N$, which in turn consists of the covariates $\mathbf{x}_i \in \mathbb{R}^d$ and the responses $y_i \in \mathbb{R}$. $\epsilon_i$ is the noise term, which is independent of the covariates. The regression function $f$ is modeled as a random function drawn from a Gaussian process (GP) prior with a zero mean function (a choice that can be relaxed if needed). A GP is a distribution over functions $f : \mathbb{R}^d \to \mathbb{R}$:
\[
f \mid \boldsymbol{\zeta} \sim \mathcal{GP}(0, k(\cdot, \cdot; \boldsymbol{\zeta})),
\]
where $k$ is a positive definite kernel function depending on hyperparameters $\boldsymbol{\zeta}$. This prior implies that, for any finite set of covariates $\{\mathbf{x}_1, \dots, \mathbf{x}_N\}$, the vector of function evaluations $\mathbf{f} = (f(\mathbf{x}_1), \dots, f(\mathbf{x}_N))^\top$ follows a multivariate normal distribution. Integrating out the latent function then yields a marginal distribution for the observed response vector:
\begin{equation} 
    \mathbf{y} \mid \boldsymbol{\zeta}, \mathbf{x}, \sigma^2 \sim \mathcal{N}(0, \mathbf{K}(\mathbf{x}; \boldsymbol{\zeta}) + \sigma^2 \mathbf{I}), 
\end{equation}
where the covariance matrix $\mathbf{K}(\mathbf{x}; \boldsymbol{\zeta})$ is defined elementwise as $\mathbf{K}(\mathbf{x}; \boldsymbol{\zeta})_{ij} = k(\mathbf{x}_i, \mathbf{x}_j; \boldsymbol{\zeta})$ \citep{rasmussen2005gaussian}, and $\mathbf{I}$ is the $N \times N$ identity matrix.

\begin{figure}[t]
    \centering
    \includegraphics[width=0.95\textwidth]{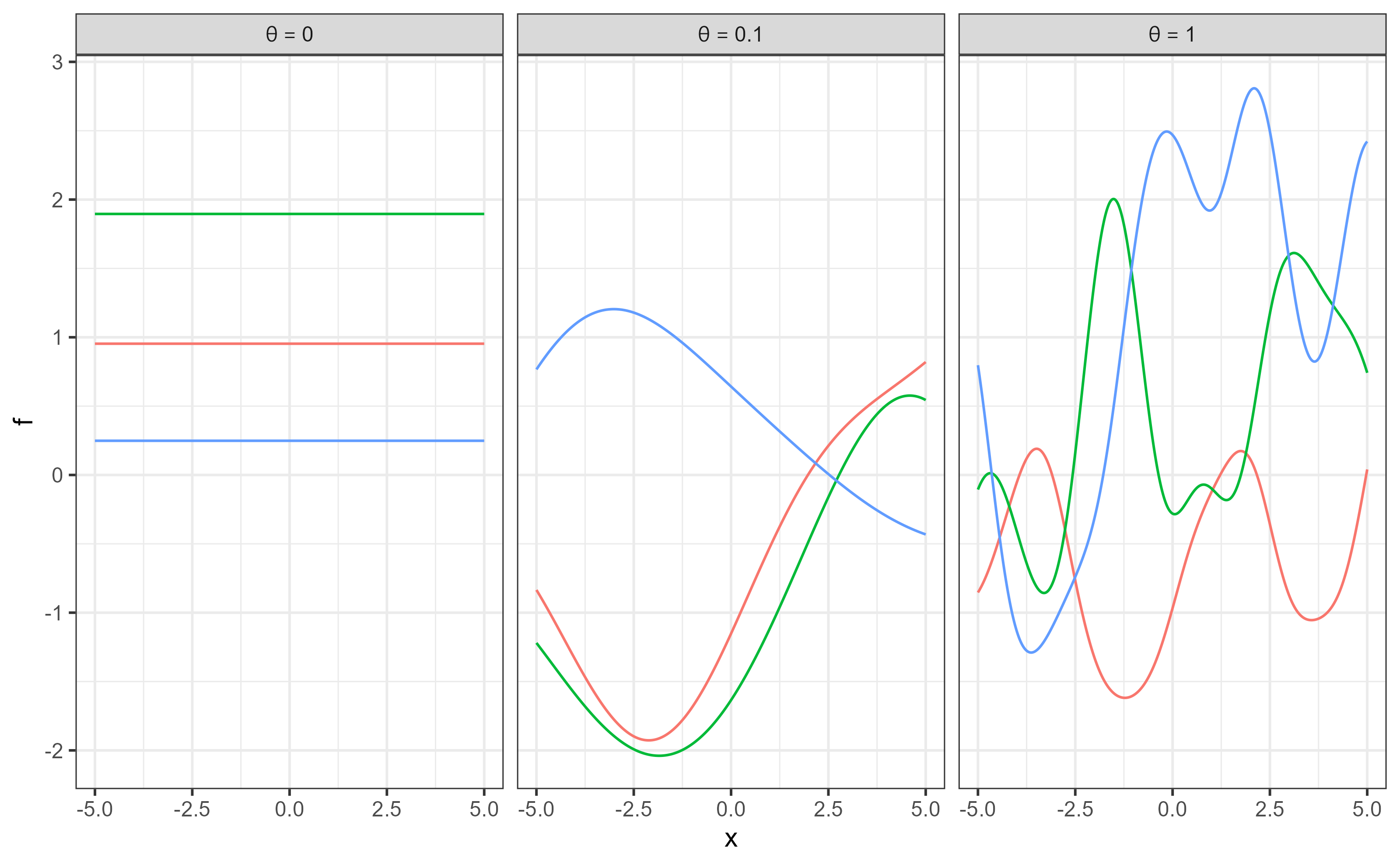}
    \caption{Samples from a GP with a squared exponential kernel, with varying values of $\theta_j$. As $\theta_j$ becomes smaller, the draws become smooother, with the extreme case of $\theta_j = 0$ leading to a constant function.}
    \label{fig:shri_effect}
\end{figure}

Many choices are feasible for the covariance function $k(\cdot, \cdot; \boldsymbol{\zeta})$. Examples include the squared exponential kernel, the Matérn kernel, the rational quadratic kernel, or the periodic kernel. While these differ in characteristics like implied smoothness, correlation between "distant" observations, or periodicity, many of them (when used in the context of regression) have at their core an anisotropic distance function of the form
\begin{equation} \label{eq:dist}
        \delta(\mathbf{z}, \mathbf{z}'; \boldsymbol{\theta}) = \sqrt{\sum_{j=1}^d (z_j - z_j')^2 \theta_j},
\end{equation}
where $\theta_j$ determines the contribution of $j$-th covariate to the distance. Consider, as an example, the squared exponential kernel written in this way:
\begin{equation}
    k_{\text{SE}}(\mathbf{z}, \mathbf{z}'; \tau, \boldsymbol \theta) = \frac{1}{\tau} \exp\left(-\frac{1}{2}\delta(\mathbf{z}, \mathbf{z}'; \boldsymbol{\theta})^2\right),
\end{equation}
where $\tau$ determines the overall variance of the GP. In this case, the hyperparameters that control the covariance structure are $\boldsymbol \zeta = \{\boldsymbol{\theta}, \tau\}$. The parameters contained in $\boldsymbol{\theta}$ dictate how each covariate contributes to the distance function, in turn directly influencing the covariance structure of the GP. As $\theta_j$ becomes small, the $j$-th covariate contributes less to the distance function, resulting in reduced influence on the covariance structure. In the limit, when $\theta_j = 0$, the $j$-th covariate no longer affects the covariance structure and can be considered irrelevant. This effect is clearly illustrated in Figure~\ref{fig:shri_effect}, which shows realizations from a GP with varying $\theta_j$ values. Smaller values of $\theta_j$ result in smoother functions, while $\theta_j = 0$ leads to a constant function, where the covariate has no more influence on the outcome. Notably, the covariance structure is governed by only a handful of hyperparameters, which simplifies the estimation process discussed in Section~\ref{sec:estim}. Despite this simplicity, the GP retains high flexibility, making it suitable for modeling complex relationships.
% Place figure as close as possible and at top of page

When $d$ is large, i.e., when one is faced with many potential predictors, a mechanism is required to remove irrelevant covariates from the model. In such high-dimensional settings, a reasonable assumption is that many covariates are irrelevant, contributing noise rather than predictive value. In the more traditional linear regression setting, this is a well-studied problem, with a popular Bayesian approach being global-local shrinkage priors \citep{polson2011shrink}:
\begin{equation}
    \beta_j \mid \tau, \lambda_j \sim \mathcal{N}(0, \tau \lambda_j), \quad \lambda_j \sim \pi_\lambda(\lambda_j), \quad \tau \sim \pi_\tau(\tau),
\end{equation}
where $\beta_j$ is the regression coefficient for the $j$-th covariate, $\lambda_j$ is the local shrinkage parameter, and $\tau$ is the global shrinkage parameter. The idea behind these priors is that $\tau$ pulls all $\beta_j$ towards zero, while local parameters $\lambda_j$ allow select $\beta_j$ to remain non-zero if supported by the data. Such priors are hierarchical in nature, with the global parameter $\tau$ inducing a joint distribution that shares information across coefficients and enables adaptation to varying sparsity levels. Various choices of $\pi_\lambda(\lambda_j)$ and $\pi_\tau(\tau)$ have been studied in the literature, such as the normal-gamma prior \citep{gri-bro:inf}, the Bayesian Lasso \citep{par-cas:bay}, the horseshoe prior \citep{carvalho2010horseshoe} or the triple gamma prior \citep{cad-etal:tri}, to name a few.

In their default specification, global-local shrinkage priors are not directly applicable to the GPR framework, as they are defined for regression coefficients with support on $\mathbb{R}$. In contrast, the parameters that govern the covariance structure in GPR, $\theta_1, \dots, \theta_d$, must remain positive to ensure a valid distance function and meaningful covariance structure. To address this, this work adopts the approach of \cite{fruhwirth2010stochastic}, which is well-established in the time-varying parameter literature, and sets $\theta_j = \beta_j^2$. This transformation replaces the normal distribution at the lowest level of the hierarchy with a gamma distribution:
\begin{equation} \label{eq:gl_pos}
    \theta_j \mid \tau, \lambda_j  \sim \mathcal{G}\left(\frac 1 2, \frac{1} {2\tau \lambda_j}\right), \quad \lambda_j \sim \pi_\lambda(\lambda_j), \quad \tau \sim \pi_\tau(\tau),
\end{equation}
which defines a class of shrinkage priors with support on $\mathbb{R}^+$, while retaining the global-local structure, enabling shared shrinkage across parameters via $\tau$ and flexibility for individual $\theta_j$ through $\lambda_j$.

This work specifically considers the hierarchical triple gamma prior of \cite{cad-etal:tri} for $\theta_j$, which is well-suited for high-dimensional GPR due to its flexibility, generality, and computational tractability. While the triple gamma prior is traditionally represented as a compound distribution of three gamma random variables (hence the name), its global-local form closest to \eqref{eq:gl_pos} is given by
\begin{equation} \label{eq:tg}
    \theta_j \mid \tau, \lambda_j \sim \mathcal{G}\left(\frac 1 2, \frac{1} {2\tau \lambda_j}\right), \quad \lambda_j \mid a, c \sim F\left(2a, 2c\right), \quad  \tau \mid a, c \sim F\left(2c, 2a\right),
\end{equation}
where $F(d_1, d_2)$ is the F distribution with degrees of freedom $d_1$ and $d_2$. The triple gamma prior offers three key advantages. First, it is a generic choice of shrinkage prior, encompassing many well-known priors as special cases (e.g., the other global-local shrinkage priors discussed previously).  Second, it is highly flexible, with the parameters $a$ and $c$ modifying the pole and tail behavior, respectively. Specifically, the pole around the origin becomes more pronounced as $a$ decreases, while the tails of the distribution become heavier as $c$ decreases. The pronounced pole effectively squelches noise, while the heavy tails allow signals to filter through. Finally, and importantly for the estimation procedure presented in Section~\ref{sec:estim}, the prior on $\theta_j$ under the triple gamma marginalised with respect to $\lambda_j$ is available in closed form:
\begin{equation} \label{tg_def}
    f(\theta_j; a, c, \tau) = \frac{\Gamfun{c+{\frac{1}{2}}}}{\sqrt{2\pi \kappa \theta_j} \Betafun{a,c}}
		\Uhyp{c + { \frac{1}{2}}, \frac{3}{2} - a, \frac{\theta_j}{2 \kappa} },
\end{equation}
where $\kappa = \tau\frac{c}{a}$ and $\Uhyp{a,b,z}$ is the confluent hypergeometric function of the second kind:
\begin{align*}
\Uhyp{a,b,z} = \frac{1}{\Gamma(a)}\int_0^\infty e^{-zt} t^{a-1}(1+t)^{b-a-1} dt.
\end{align*}
The closed-form marginal density simplifies the estimation procedure by collapsing the full hierarchical structure of the prior, keeping the number of hyperparameters to   estimate per individual $\theta_j$ low. This computational efficiency, combined with the prior's flexibility and generality, makes it a strong choice for the GPR framework presented here.

\subsection{The Importance of Hierarchy}

GPR models suffer from the curse of dimensionality in a specific way. Consider two points in the output space, $y$ and $y'$. When a new covariate is added, the anisotropic distance in \eqref{eq:dist} can only increase\footnote{Technically, it could remain constant if the associated $\theta_j = 0$. However, given the use of continuous shrinkage priors, this will happen with probability $0$.}, leading to reduced covariance between $y$ and $y'$, and, more broadly, to an increasingly sparse covariance matrix. This sparsification occurs mechanically as the number of covariates grows, even if they do not influence the outcome of interest \citep{binois2022survey}. This sparsification not only affects the covariance structure but can also lead to overfitting or reduced predictive accuracy, particularly in high-dimensional settings.

The priors on $\theta_1, \dots, \theta_d$ are critical in mitigating the curse of dimensionality by pushing the $\theta_j$ values of irrelevant covariates toward zero, effectively excluding them. The choice of prior for the global shrinkage parameter $\tau$ plays a particularly important role here. As shown in \cite{cad-etal:tri}, under the triple gamma shrinkage prior, setting $\tau \sim F(2c, 2a)$ implies a uniform prior on the \textit{model size}, i.e., the number of relevant covariates. This ensures that some prior mass remains on models with few or no covariates, even when $d$ is large, while still allowing flexibility for larger models when supported by the data.

To be more precise on what "model size" means, first consider the following equivalent re-parameterization of the triple gamma prior \citep{cad-etal:tri}:
\begin{equation}
    \theta_j \mid \rho_j \sim \mathcal{G}\left(\frac 1 2, \frac{\rho_j} {2(1 - \rho_j)}\right), \quad \rho_j \mid a, c, \tau \sim \mathcal{TPB}\left(a, c, \frac{2c}{\tau a}\right),
\end{equation} 
where $\mathcal{TPB}$ is the three-parameter beta distribution. Importantly, $\rho_j \in [0, 1]$ and is called the \textit{shrinkage factor}, as it characterizes the amount of shrinkage applied to $\theta_j$. 

This formulation generalizes the concept of shrinkage factors, well-studied in normal-means and regression settings \citep{carvalho2010horseshoe}, to positive-constrained parameters. The interpretation remains straightforward: as $\rho_j$ approaches zero, $\frac{\rho_j} {2(1 - \rho_j)}$ approaches zero, making the prior for $\theta_j$ converge to a gamma distribution with infinite mean and variance. Conversely, as $\rho_j$ approaches one, $\frac{\rho_j} {2(1 - \rho_j)}$ approaches infinity, driving the prior for $\theta_j$ toward a gamma distribution with mean and variance both zero. Thus, $\rho_j = 1$ corresponds to total shrinkage, while $\rho_j = 0$ implies no regularization on the parameter. Despite its origins in the sparse normal-means problem, this generalization of shrinkage factors adapts naturally to the square-transform case.

The shrinkage factor naturally leads to the concept of model size $K$, defined as the number of $\theta_j$ where $\rho_j < 0.5$: 
\begin{equation}
    K = \sum_{j=1}^d \mathbb{I}(\rho_j < 0.5).
\end{equation}
$K$ corresponds to the number of covariates that significantly contribute to the covariance matrix and is itself a random variable, with a distribution implied by the choice of hierarchical shrinkage prior:
\begin{equation}
    K \sim \text{Binom}(d, \pi), \quad \pi = \Pr(\rho_j < 0.5).
\end{equation}
What \cite{cad-etal:tri} show is that under $\tau \sim F(2c, 2a)$, the induced probability $\pi = \Pr(\rho_j < 0.5)$ is uniformly distributed, implying a uniform prior on the model size. This is in stark contrast to a non-hierarchical prior, where $\pi$ is some fixed constant, a choice that is highly informative about the model size. The uniform prior ensures that models with few covariates are not overly penalized, which is critical for achieving robust performance in high-dimensional settings where irrelevant covariates dominate.

\begin{figure}[t]
    \centering 
    \includegraphics[width=0.95\textwidth]{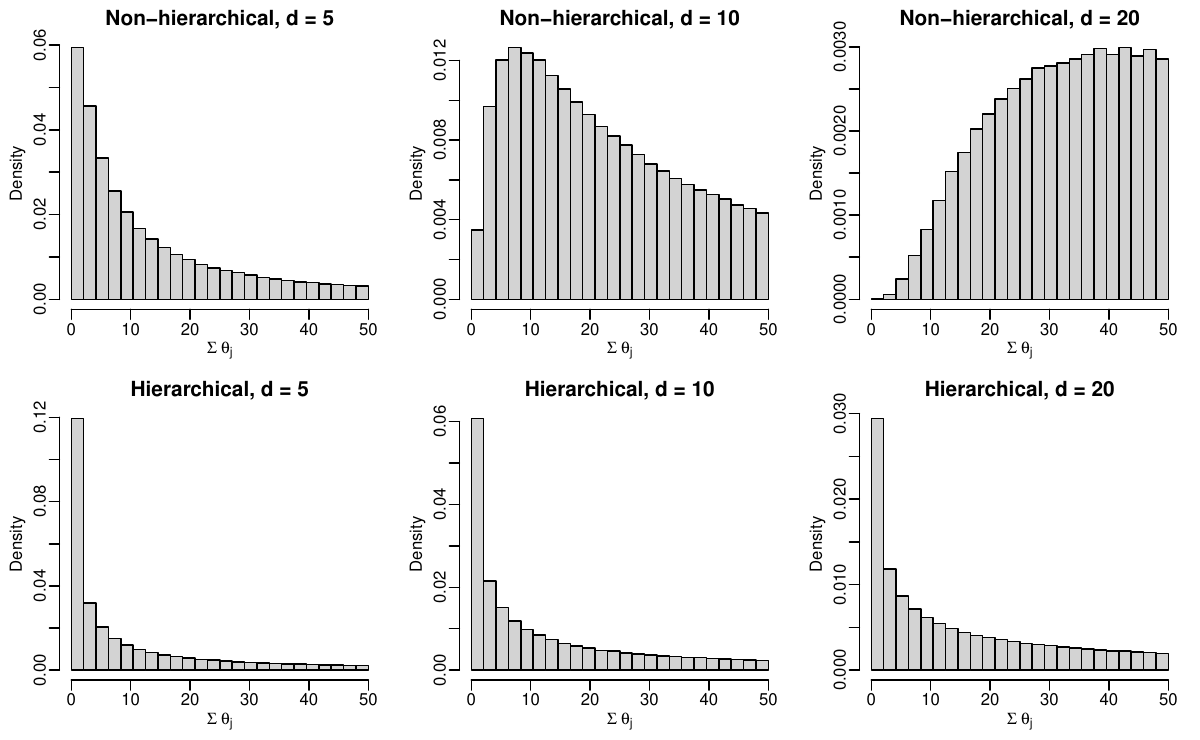}
    \caption{Samples from $\sum_{j=1}^d\theta_j$ under the hierarchical triple gamma prior (bottom) and the non-hierarchical triple gamma prior (top), for different $d$. The hierarchical prior places more mass on smaller values of $\sum_{j=1}^d\theta_j$, even as the number of covariates increases.}
    \label{fig:sum_effect}
\end{figure}

The implied $\sum_{j=1}^d\theta_j$ can serve as a measure of the \textit{a priori} distance between observations encoded within the GP prior. Figure~\ref{fig:sum_effect} shows the distribution of this measure across various $d$ for both the hierarchical and non-hierarchical triple gamma prior. The hierarchical prior concentrates more mass on smaller values of $\sum_{j=1}^d\theta_j$, even as the covariate space increases. In contrast, the mode of the distribution under the non-hierarchical prior increases with the number of covariates, despite having the same conditional distribution. This behavior arises because the hierarchical prior induces a joint distribution over $\theta_1, \dots, \theta_d$, where any large $\theta_j$ decreases the probability of other $\theta_{-j}$ values being large.

\section{Estimation via Normalizing Flows} \label{sec:estim}

Estimation of GPR models is a notoriously difficult problem, despite the log marginal likelihood (here marginalized over the function $f(\mathbf{x})$) being available in closed form:
\begin{equation}
    \label{eq:loglik}
\begin{aligned} 
\log p\left(\mathbf y \mid \tau, \theta_1, \dots, \theta_d, \sigma^2\right) =&-\frac{N}{2} \log (2 \pi)-\frac{1}{2} \log \left|\mathbf{K}(\mathbf{x}; \boldsymbol{\theta}, \tau)+\sigma^2 I\right| \\
&-\frac{1}{2} \mathbf y^{T}\left(\mathbf{K}(\mathbf{x}; \boldsymbol{\theta}, \tau)+\sigma^2 I\right)^{-1} \mathbf y. \nonumber
\end{aligned}
\end{equation}

Two main problems arise: First, the triple gamma prior is not (conditionally or otherwise) conjugate to the posterior of the parameters, which precludes any closed form solutions and makes sampling based approaches less attractive. Second, the marginal likelihood is expensive to evaluate, as it requires the inversion the $N \times N$ matrix $\mathbf{K}(\mathbf{x}; \boldsymbol{\theta}, \tau) + \sigma^2 I$, which has a computational cost of order $\mathcal O(N^3)$. This problem is compounded in the fully Bayesian approach, as sampling based methods often require the marginal likelihood to be evaluated at each iteration. Many widely used variational methods, such as the mean-field approximation, alleviate the computational burden but often lack expressivity, leading to poor approximations of the posterior distribution.

This work proposes a different approach to the estimation of GPR models under hierarchical shrinkage priors. Specifically, it explores the use of normalizing flows \citep{rezende2015variational} to approximate the posterior distribution in a variational inference framework. While still an approximation to the true posterior, normalizing flows are highly flexible and can approximate even complex posterior shapes in a computationally tractable way. This makes them a step up from widely used mean-field approximations, which, by design, are unable to capture correlations between parameters. This may be of particular importance in a regression setting, where explanatory variables are often correlated. 

The key computational advantage of normalizing flows is that the method can easily take advantage of parallelized computing available via Graphics Processing Units (GPUs). This means that a rich approximation of the posterior can be obtained, while still remaining computationally tractable. This is in contrast to sampling based approaches and other variational inference methods, where the former is inherently sequential and the latter often lacks the flexibility to capture complex posterior shapes. 

\subsection{A Brief Introduction to Normalizing Flows}

Normalizing flows are a class of flexible, invertible transformations that approximate complex distributions by mapping a simple base distribution onto a more complex target distribution through a series of simpler transformations. This makes them an effective tool for variational inference, where capturing complex posterior shapes is often challenging.

To illustrate, consider the goal of defining a joint distribution over a $D$-dimensional random variable $\mathbf{Z}$.\footnote{Note that, in this application, the $\mathbf{Z}$ we wish to approximate is ultimately the posterior distribution.} The key idea behind normalizing flows is to introduce a simple base distribution $p_{\mathbf{U}}(\mathbf{U})$ over a random variable $\mathbf{U}$ and then express $\mathbf{Z}$ as a transformation $T$ of $\mathbf{U}$:
\[
\mathbf{Z} = T(\mathbf{U}), \quad \mathbf{U} \sim p_{\mathbf{U}}(\mathbf{U}).
\]
To be of interest for variational inference, the transformation $T$ depends on parameters, denoted here as $\boldsymbol{\phi}$, leading to a distribution $p_{\mathbf{Z}}(\mathbf{Z}; \boldsymbol{\phi})$ over $\mathbf{Z}$\footnote{Technically, one could assume that the base distribution $p_{\mathbf{U}}(\mathbf{U})$ also depends on parameters. However, as will be seen later, one can often subsume the parameters of the base distribution into the transformation $T$, so they are omitted here.}. Out of notational convenience, we will often drop the dependence on these parameters and write, e.g., $p_{\mathbf{Z}}(\mathbf{Z})$ instead. For $T$ to qualify as a valid flow, it must be a diffeomorphism, meaning it is invertible and both $T$ and its inverse $T^{-1}$ are differentiable. This constraint also requires that $\mathbf{U}$ is of the same dimension $D$ as $\mathbf{Z}$. Under these conditions, the density of $\mathbf{Z}$ can be computed using the change of variables formula:
\[
p_{\mathbf{Z}}(\mathbf{Z}) = p_{\mathbf{U}}(\mathbf{U}) \left| \operatorname{det} J_T(\mathbf{U}) \right|^{-1},
\]
where $\mathbf{U} = T^{-1}(\mathbf{Z})$ and $J_T(\mathbf{U})$ is the Jacobian matrix of $T$ evaluated at $\mathbf{U}$.

Now, consider a sequence of transformations, $T_1, \dots, T_K$, and define $T$ as their composition, $T = T_K \circ \ldots \circ T_1$. The resulting density of $\mathbf{Z}$ is then:
\[
p_{\mathbf{Z}}(\mathbf{Z}) = p_{\mathbf{U}}(\mathbf{U}) \prod_{k=1}^K \left| \det J_{T_k}(\mathbf{U}_k) \right|^{-1},
\]
where each $T_k$ is a transformation with parameters $\boldsymbol{\phi}_k$ and $\mathbf{U}_k$ represents the intermediate transformed variable after applying $T_k$. While each individual transformation $T_k$ may be simple, the composition of them is able to approximate more complex distributions.  This structure is akin to a neural network for density estimation, with each transformation acting as a "layer" subject to constraints that ensure invertibility and differentiability.

\subsubsection{An Example Transformation: The Radial Flow}

There exist many possible transformations that could act as individual layers in the flow, such as the planar and radial flows from the seminal paper of \cite{rezende2015variational}, the Inverse Autoregressive Flow (IAF) of \cite{kingma2016improved}, or the Sylvester normalizing flow of \cite{berg2018sylvester}, to name a few. While they differ in their specific form, a key property they all share is that they are designed such that the Jacobian and its determinant can be computed efficiently. This is mostly achieved through the use of triangular Jacobian matrices, which have a determinant that can be computed in $\mathcal O(D)$ time, where $D$ is the dimension of the random variable one wishes to model. 

\begin{figure}[t]
    \centering
    \includegraphics[width=0.95\textwidth]{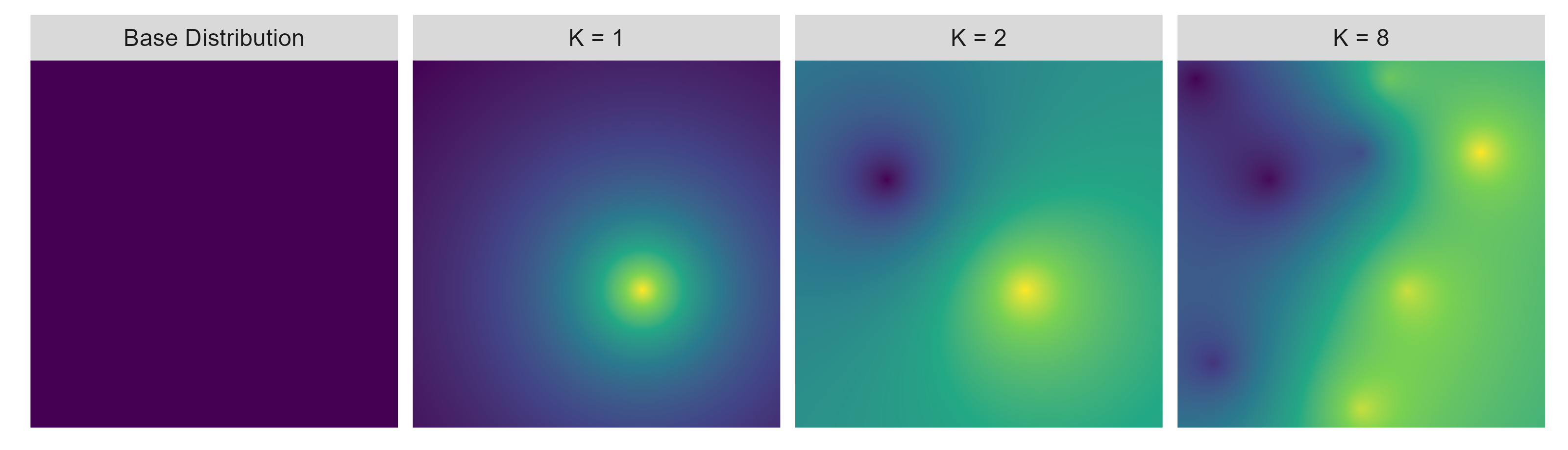}
    \caption{Effect of radial normalizing flows on a 2D uniform distribution. The base distribution (left) is progressively transformed by radial flows, introducing localized contractions or expansions around specified centers, controlled by $\alpha$ (spread) and $\beta$ (intensity). Density values are normalized to the range $[0, 1]$, with higher densities in yellow and lower densities in purple.}

    \label{fig:radial_flow}
\end{figure}

To build intuition for normalizing flows, consider, as an example, the radial flow, which is defined as:
\begin{equation*}
    T(\mathbf{U}; \beta, \alpha, \mathbf{U}_0) = \mathbf{U} + \beta \frac {\mathbf{U} - \mathbf{U}_0} {\alpha + \|\mathbf{U} - \mathbf{U}_0\|_2},
\end{equation*}
with parameters $\beta, \alpha, \mathbf{U}_0$. The basic intuition behind this transformation is that it expands or contracts the base distribution around the point $\mathbf{U}_0$, either pushing mass towards $\mathbf{U}_0$ or pulling it away. The spread of the contraction is controlled by $\alpha$ and the direction and intensity is controlled by $\beta$. Through successive applications of this transformation, it is possible to approximate distributions that are far more complex than would appear possible at first glance. Figure~\ref{fig:radial_flow} shows the effect of radial flows on a 2D uniform distribution, with the base distribution progressively transformed by radial flows, introducing localized contractions and expansions. Note that each layer has its own parameters $\beta_k, \alpha_k, \mathbf{U}_{0,k}$, leading to different transformations at each step. These are the parameters that can be learned during estimation, with the goal of approximating the posterior distribution as closely as possible.

\subsubsection{Normalizing Flow used in this Work}

While the transformation behind the radial flow is simple and therefore good for building intuition, it suffers in terms of expressivity. As it can only contract or expand the base distribution around a single point, the number of layers $K$ needs to be large to make the flow expressive \citep{papamakarios2021normalizing}. As such, more complex transformations are often used in practice, as in this work, where the Sylvester normalizing flow is employed \citep{berg2018sylvester}:
\begin{equation*}
    T(\mathbf{U}; \mathbf{Q}, \mathbf{W}, \mathbf{b}) = \mathbf{U}+ \mathbf{Q}\mathbf h(\mathbf W^T \mathbf{U}+ \mathbf b),    
\end{equation*}
where $\mathbf{U}\in \mathbb R^D$, $\mathbf{Q}, \mathbf W \in \mathbb R^{D \times D}$, $\mathbf b \in \mathbb R^D$ and $\mathbf h$ is an elementwise non-linear function (usually the hyperbolic tangent). It is a generalization of the planar flow of \cite{rezende2015variational} that increases the expressivity of each individual transformation. The increased expressivity of each individual layer means that fewer layers overall are required to reach the same level of overall expressivity.

Ensuring the transformation is invertible requires restricting the parameter space for $\mathbf{Q}$, $\mathbf{W}$, and $\mathbf{b}$; see \cite{berg2018sylvester} for details. It is important to note that the proposed methodology is not tied to Sylvester normalizing flows and can accommodate other transformations, if they are better suited to a given problem at hand.

\subsection{Normalizing Flows for Variational Inference}

Normalizing flows' capability to approximate complex distributions in a black-box fashion makes them well-suited for variational inference. If we let $\boldsymbol \xi$ contain all the parameters of a model we wish to estimate, then the objective of variational inference is to approximate the posterior distribution $p(\boldsymbol{\xi} \mid \mathbf{y})$ by selecting the best approximation from a family of distributions $q(\boldsymbol{\xi}; \boldsymbol{\phi})$, which in turn depends on the parameters $\boldsymbol{\phi}$. This is achieved by minimizing the Kullback-Leibler (KL) divergence between the approximate distribution $q(\boldsymbol{\xi}; \boldsymbol{\phi})$ and the true posterior $p(\boldsymbol{\xi} \mid \mathbf{y})$:
\[
q^*(\boldsymbol{\xi}; \boldsymbol{\phi}) = \operatorname{arg\,min}_{\boldsymbol{\phi}} \, \text{KL}(q(\boldsymbol{\xi}; \boldsymbol{\phi}) \mid\mid p(\boldsymbol{\xi} \mid \mathbf{y})).
\]
The key to using normalizing flows for variational inference is to let the family of approximating distributions $q(\boldsymbol{\xi}; \boldsymbol{\phi})$ be defined as a normalizing flow with $K$ transformations:
\[
q(\boldsymbol{\xi}; \boldsymbol{\phi}) = p_{\mathbf{U}}(\mathbf{U}) \prod_{k=1}^K \left| \operatorname{det} J_{T_k}(\mathbf{U}_k) \right|^{-1}.
\]
where $\boldsymbol \phi$ now contains the parameters of all transformations $T_1, \ldots, T_K$.  Note that this is in contrast to the mean-field approximation, which assumes a factorized posterior with independent components, resulting in approximations that cannot capture dependencies. Normalizing flows, on the other hand, naturally capture dependencies within the posterior, as the transformations $T_k$ explicitly induce dependencies between the parameters.

A common approach to minimizing the KL-divergence is to maximize the Evidence Lower BOund (ELBO). When using normalizing flows as the approximate posterior, an ELBO of the following form arises \citep{papamakarios2021normalizing}:
\begin{equation} \label{eq:elbo}
\text{ELBO}(\boldsymbol{\phi}) = \mathbb{E}_{p_{\mathbf{U}}(\mathbf{U})}[\log p(\mathbf{y}, T(\mathbf{U}; \boldsymbol{\phi}))] + \mathbb{E}_{p_{\mathbf{U}}(\mathbf{U})}\left[\log \left| \operatorname{det} J_T(\mathbf{U}; \boldsymbol{\phi}) \right|\right].
\end{equation}
The reason why this approach to estimation is attractive, is that this expression can be optimized via stochastic gradient descent, which is computationally efficient and can, importantly, be run on GPUs, greatly speeding up computation. 

\subsubsection{Practical Considerations}

\begin{algorithm}[ht!]
  \spacingset{1.0}
    \caption{Normalizing Flows for Variational Inference in GPR Models}\label{alg:nf_gpr}
    \begin{algorithmic}[1]
    \Require Training data $\mathcal{D} = \{(\mathbf{x}_i, y_i)\}_{i=1}^n$, kernel function $k(\cdot, \cdot; \boldsymbol{\theta}, \tau)$, base distribution $p_{\mathbf{U}}(\mathbf{U})$, sequence of flow functions $\{T_k(\cdot; \boldsymbol{\phi}_k)\}$, number of samples $S$ from base distribution, number of iterations $M$
    \State Initialize variational parameters $\boldsymbol{\phi}_k$ for each flow layer \( k = 1, \ldots, K \)   
    \For{$i = 1$ to $M$}
        \State Sample \( S \) realizations \( \{\mathbf{u}_0^{(s)}\}_{s=1}^S \) from the base distribution \( p_{\mathbf{U}}(\mathbf{U}) \)
        \For{$s = 1$ to $S$}
            \State Initialize \( \log J^{(s)} \gets 0 \)
            \For{$k = 1$ to $K$}
                \State Apply flow transformation: \( \mathbf{u}_k^{(s)} \gets T_k(\mathbf{u}_{k-1}^{(s)}; \boldsymbol{\phi}_k) \)
                \State Update cumulative log det.: 
                \(
                \log J^{(s)} \gets \log J^{(s)} + \log \left| \det J_{T_k}\left(\mathbf{u}_{k-1}^{(s)}; \boldsymbol{\phi}_k \right) \right|
                \)
            \EndFor
            \State Apply softplus transformation: \( \boldsymbol{\theta}^{(s)} \gets \text{softplus}(\mathbf{u}_K^{(s)}) \)
            \State Update cumulative log det.:
            \(
            \log J^{(s)} \gets \log J^{(s)} + \log \left| \det J_{\text{softplus}}\left(\mathbf{u}_K^{(s)} \right) \right|
            \)
        \EndFor
        \State Approximate ELBO using Monte Carlo: 
        \[
        \text{ELBO}_{\text{MC}}(\boldsymbol{\phi}) \gets \frac{1}{S} \sum_{s=1}^S \left[ \log p\left(\mathbf{y}, \boldsymbol{\theta}^{(s)} \right) 
        + \log J^{(s)} \right]
        \]
        \State Compute gradients \( \nabla_{\boldsymbol{\phi}} \text{ELBO}_{\text{MC}}(\boldsymbol{\phi}) \)
        \State Update \( \boldsymbol{\phi} \) using a gradient ascent method
    \EndFor
    \State Return optimized variational parameters \( \boldsymbol{\phi} \)
    \end{algorithmic}
\end{algorithm}

While the overview given so far contains all the theory to implement normalizing flows for GPR models, a few extra comments on practical considerations may make the procedure somewhat more approachable. First, the ELBO presented in \eqref{eq:elbo} is difficult to compute directly, as it requires the evaluation of the expectation over the base distribution $p_{\mathbf{u}}(\mathbf{u})$. Therefore, in practice, the ELBO is approximated using Monte Carlo (MC) sampling. To this end, $S$ samples $\{\mathbf{u}_0^{(s)}\}_{s=1}^S$ are drawn from the base distribution $p_{\mathbf{u}}(\mathbf{u})$ and transformed succesively by the $K$ flow functions to obtain $\{\mathbf{u}_K^{(s)}\}_{s=1}^S$, as well as the associated Jacobians $\{\log J^{(s)}\}_{s=1}^S$. Further, as the hyperparameters of a GPR model are all constrained to be positive, a softplus transformation $g(x) = 1/\beta \log(1 + \exp(\beta x))$ is applied to $\mathbf{u}_K^{(s)}$ to ensure that the parameters remain positive. This gives the current samples from the posterior approximation used for further calculations $\boldsymbol\xi^{(s)} = g\left(\mathbf{u}_K^{(s)}\right)$. Importantly, as this is also a transformation, the Jacobian of the softplus must be accounted for in the ELBO. The ELBO can then be approximated as:
\[
\text{ELBO}_{\text{MC}}(\boldsymbol{\phi}) = \frac{1}{S} \sum_{s=1}^S \left[ \log p\left(\mathbf{y}, \boldsymbol{\xi}^{(s)} \right) + \log J^{(s)} \right].
\]
The unnormalized log posterior $\log p\left(\mathbf{y}, \boldsymbol{\xi}^{(s)}\right)$ evaluated at the MC samples then reads:
\vspace{-1em}
    \begin{align}
        \log p\left(\mathbf{y}, \boldsymbol{\xi}^{(s)}\right) &= \log p\left(\mathbf{y} \mid \boldsymbol{\xi}^{(s)} \right) + \log p\left(\boldsymbol{\xi}^{(s)} \right) \notag \\
        &= -\frac{1}{2} \log \left|\mathbf{K}(\mathbf{x}; \boldsymbol{\theta}, \tau)+\sigma^2 I\right| -\frac{1}{2} \mathbf y^{T}\left(\mathbf{K}(\mathbf{x}; \boldsymbol{\theta}, \tau)+\sigma^2 I\right)^{-1} \mathbf y \label{comp:loglik}\\
        &+ \sum_{j=1}^d \frac 1 2 \left(\log(\tau) - \log(\theta_j) \right)  + \log\left(\Uhyp{c + { \frac{1}{2}}, \frac{3}{2} - a, \frac{\theta_j}{2 \kappa} }\right) \label{comp:tg} \\
        &+ (c - 1) \log(\tau) - (c + a) \log\left(1 + \frac c a \tau\right) \label{comp:f} \\
        &- \lambda \sigma^2 \label{comp:sig} \\
        &+ C, \label{comp:const}
    \end{align}
where \eqref{comp:loglik} is the unnormalized version of the log marginal likelihood in \eqref{eq:loglik}, \eqref{comp:tg} is a sum over unnormalized log densities of the triple gamma distribution in \eqref{tg_def}, \eqref{comp:f} is the unnormalized log density of the F distribution, \eqref{comp:sig} is the unnormalized log prior on $\sigma^2$, in this case assumed to be an exponential distribution, and \eqref{comp:const} is a constant that does not depend on the parameters.

From here, the gradients of the ELBO with respect to the parameters $\boldsymbol{\phi}$ can be computed via backpropagation, and the parameters can be updated using a gradient ascent method \citep[e.g., through the pytorch library;][]{paszke2017automatic}. The entire procedure is summarized in Algorithm~\ref{alg:nf_gpr}. The loop over the samples $S$ is written out for clarity, but, in practice, can be exectued in parallel, which makes the method highly efficient and well-suited for GPU computation.

\begin{figure}[t!]
    \centering
    \includegraphics[width=0.95\textwidth]{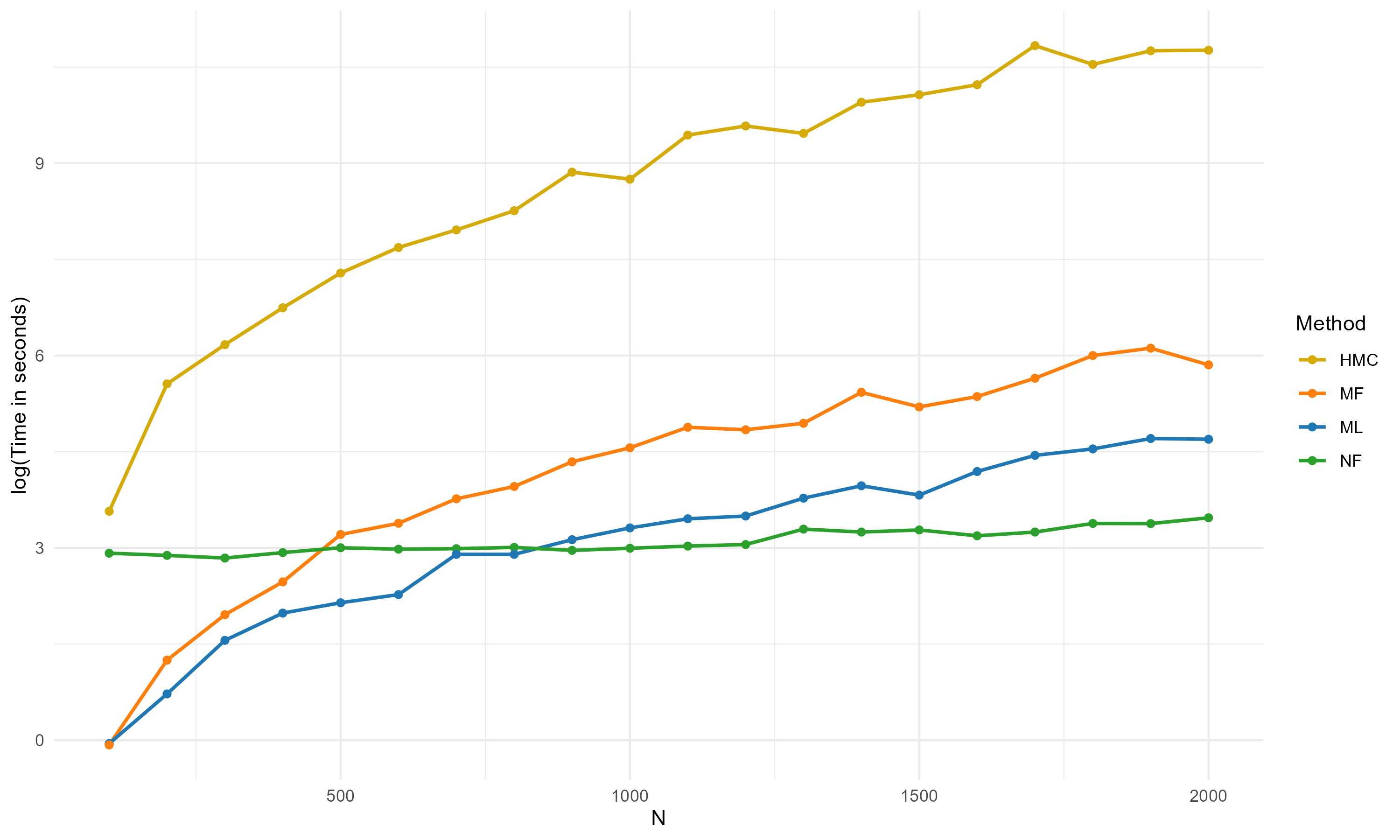}
    \caption{Log of time in seconds to compute 100 draws or iterations (depending on the method) for Hamiltonian Monte Carlo (HMC), mean-field (MF) approximation, maximum likelihood (ML) estimation, and the proposed normalizing flow (NF) approach. The number of covariates $d$ is fixed at 25.}
    \label{fig:speed_comparison}
\end{figure}

The computational efficiency of the proposed method is illustrated in Figure~\ref{fig:speed_comparison}, which shows the time required to compute 100 iterations or draws (depending on the method) for Hamiltonian Monte Carlo (HMC), mean-field variational inference (MF), maximum likelihood estimation (ML), and the proposed normalizing flow approach (NF). The number of covariates is fixed at $d = 25$. HMC and MF were implemented using the \texttt{rstan} package \citep{rstan}, ML was implemented using the \texttt{GPy} library \citep{gpy2014}, and the NF method was implemented using \texttt{torch} for \texttt{R} \citep{torchR}. All benchmarks were run on a laptop with an Intel Core i7-9750H CPU and an NVIDIA GeForce RTX 2060 GPU.

The results should not be interpreted as exact runtime comparisons, as each method is implemented in a different software framework with distinct language backends and optimization infrastructure. However, the plot provides a practical sense of the relative computational cost across approaches.

The NF method exhibits stable runtime with increasing $N$, requiring approximately 20 seconds across the entire range. While it is slower than ML and MF for small $N$, this advantage reverses at $N \approx 500$ for MF and at $N \approx 900$ for ML. This suggests that the main bottleneck in the NF implementation is likely due to fixed overhead in the underlying \texttt{torch} interface, rather than the complexity of the method itself. Once the bottleneck becomes the inversion of the covariance matrix, one would expect to see it scale similarly to the other methods. In contrast, HMC is the most computationally expensive method across all sample sizes, and becomes infeasible for larger datasets. As such, it was also excluded from the simulation study below, as it was not able to run in a reasonable time for larger datasets.

\section{Simulation Study} \label{sec:sim_study}
  
\begin{figure}[ht!]
    \centering
    \includegraphics[width=0.95\textwidth]{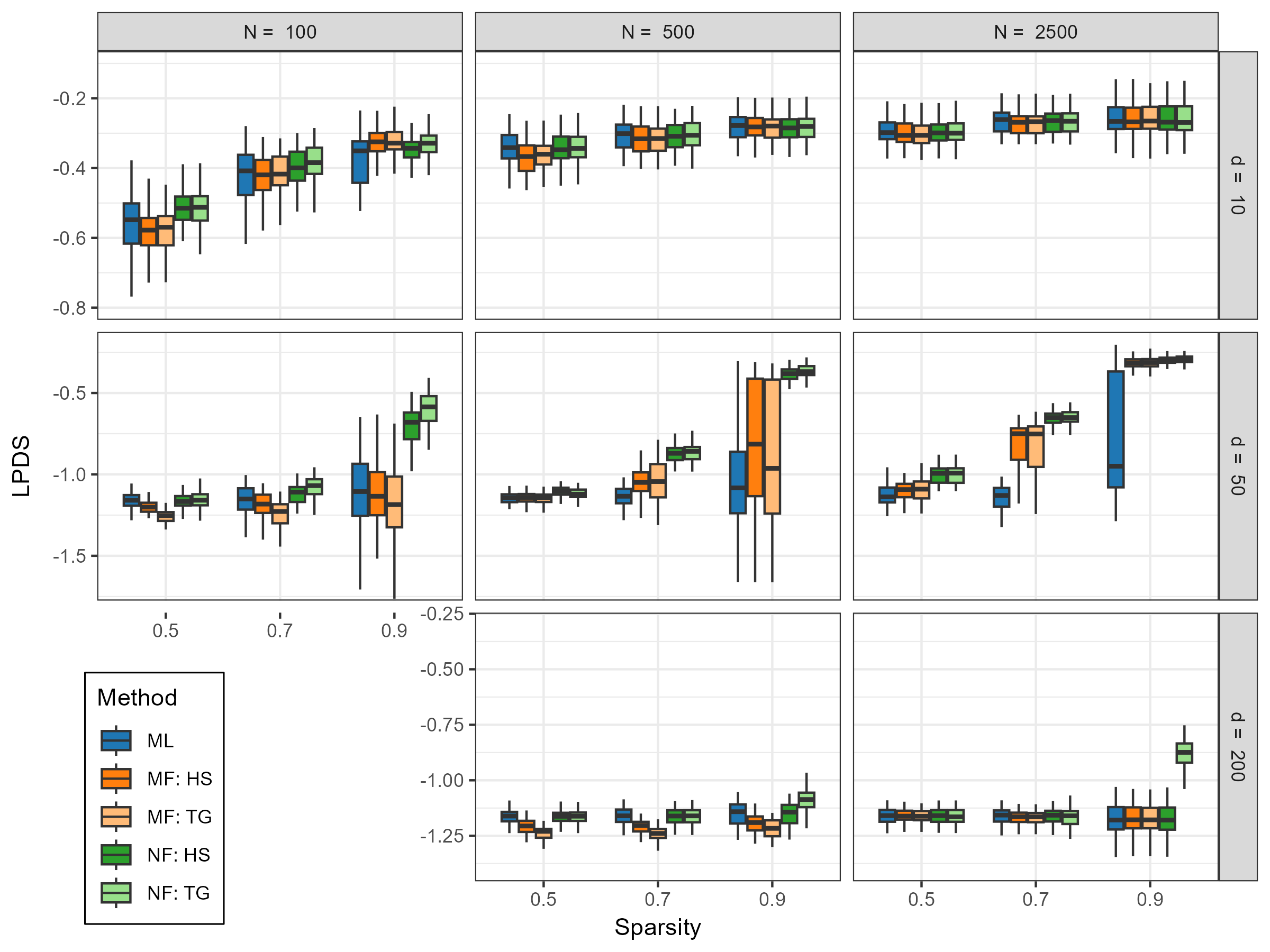}
    \caption{Log Predictive Density Scores (LPDS) for Maximum Likelihood (ML), mean-field (MF) approximation under a horseshoe (MF: HS) and triple gamma prior (MF: TG) and the proposed approach under a horseshoe prior (NF: HS) and triple gamma prior (NF: TG) with multicollinearity level $\rho = 0.5$ across varying sparsity levels, sample sizes ($N$), and dimensions ($d$). Y-axes are scaled per row.}
    \label{fig:sim_results_05}
\end{figure}

\begin{figure}[ht!]
    \centering
    \includegraphics[width=0.95\textwidth]{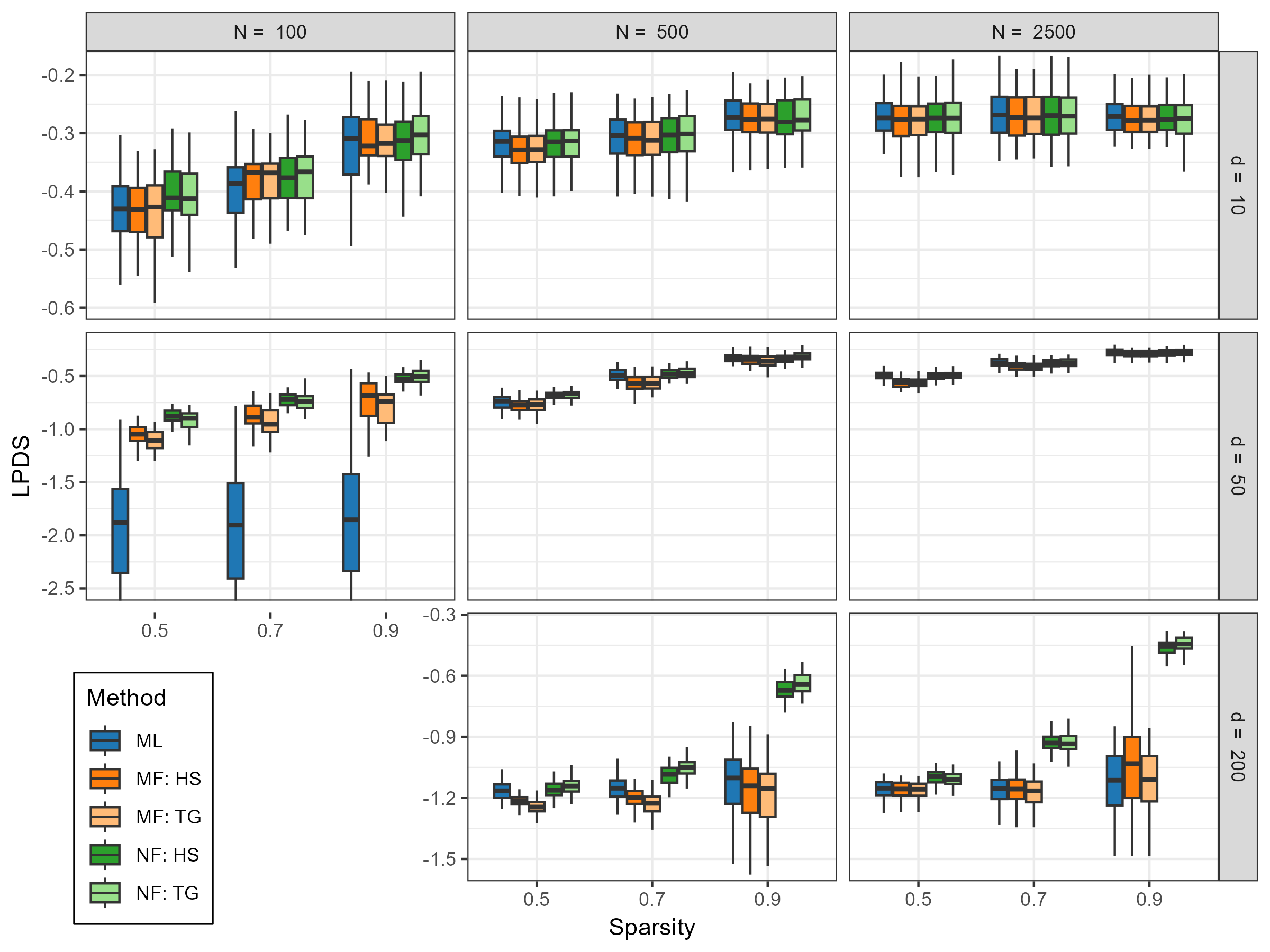}
    \caption{Log Predictive Density Scores (LPDS) for Maximum Likelihood (ML), mean-field (MF) approximation under a horseshoe (MF: HS) and triple gamma prior (MF: TG) and the proposed approach under a horseshoe prior (NF: HS) and triple gamma prior (NF: TG) with multicollinearity level $\rho = 0.9$ across varying sparsity levels, sample sizes ($N$), and dimensions ($d$). Y-axes are scaled per row.}
    \label{fig:sim_results_09}
\end{figure}

To evaluate the proposed approach to GPR, a simulation study was conducted, with a particular focus on out-of-sample predictive performance. To this end, data was simulated from a GPR model, varying the following factors:
\begin{itemize}
    \item Sparsity levels: The proportion of irrelevant covariates $s$, with $s \in \{0.5, 0.7, 0.9\}$.
    \item Dimensionality: The number of covariates $d$, with $d \in \{10, 50, 200\}$.
    \item Sample sizes: The number of observations $N$, with $N \in \{100, 500, 2500\}$.
    \item The degree of multicollinearity: The correlation between covariates, with $\rho \in \{0.5, 0.9\}$.
\end{itemize}
The covariates were generated as $\boldsymbol x_{i} \sim \mathcal{N}(\boldsymbol 0, \boldsymbol B)$, where $\boldsymbol B$ is a matrix with 1 on the main diagonal and all off-diagonal elements set to $\rho$. The inverse lengthscales $\theta_1, \dots, \theta_d$ were drawn from a gamma distribution with shape 6 and scale 24. This strikes a nice balance of ensuring the $\theta_j$ are different enough from $0$ to be detected, while not having individual $\theta_j$ being so large they dominate the covariance structure. Of the $d$ $\theta_j$, $\lfloor s \cdot d \rfloor$ were randomly set to $0$. The response variable was then generated as $\mathbf y\sim \mathcal{N}(\mathbf 0, \mathbf{K}(\mathbf{x}; \boldsymbol{\theta}, \tau) + 0.1\mathbf I)$, where $\mathbf{K}(\mathbf{x}; \boldsymbol{\theta}, \tau)$ is the covariance matrix generated by the squared exponential kernel with parameters $\boldsymbol{\theta}$ and $\tau$, which was set to 2 for all runs.

The proposed normalizing flow approach was compared to a standard GPR model with an ARD kernel, estimated using a maximum likelihood approach with the GPy library \citep{gpy2014}, as well as a mean-field variational approximation implemented in \texttt{rstan} \citep{rstan}. For the maximum likelihood method, optimization was restarted 10 times, with the best result retained. For the mean-field approximation, convergence was assessed by ensuring that the relative norm of the objective was smaller than $0.01$, the default setting in \texttt{rstan}.

For both the proposed method and the mean-field approximation, two prior specifications were considered: one with $a = c = 0.1$ and another with $a = c = 0.5$. The former represents a more aggressive triple gamma specification, with strong shrinkage near the origin and very heavy tails, while the latter corresponds to the variance shrinkage counterpart of the horseshoe prior of \cite{carvalho2010horseshoe}. In both cases, $\tau \sim F(2c, 2a)$ and $\sigma^2 \sim \text{Exp}(10)$. The approximating flows consisted of 10 layers, and 10 latent samples were used for each Monte Carlo estimate of the ELBO. To save on memory, the number of latent samples was reduced to 2 in the case of $N = 2500$. This did not seem to affect performance significantly. The normalizing flow models were run for 3000 iterations to ensure convergence, although this was typically achieved much earlier.

Each configuration was repeated across 50 independent runs. For each run, the data was split into a training set (of size $N$) and a test set of 300 observations. Performance was evaluated using the average log predictive density score (LPDS) on the test set, with higher LPDS values indicating better predictive performance.

The results are presented in Figure~\ref{fig:sim_results_05} for $\rho = 0.5$ and Figure~\ref{fig:sim_results_09} for $\rho = 0.9$. Overall, the NF approach seems to have the largest advantage when there are many covariates as well as a high degree of sparsity, as can be seen in the second and third row of the two figures. In such cases, the combination of aggressive shrinkage and rich posterior approximation of the normalizing flows seems to provide the largest benefit. While the prior is the same in the mean-field approach, the rougher mean-field approximation can not take advantage of this benefit to the same degree. Maximum likelihood performs well when the ratio of covariates to observations is not too unfavorable, but falls apart when the covariate space becomes too high-dimensional.

Zooming in on the results, the first rows show that all methods perform similarly when there are only 10 covariates, with only slight dips in performance from the mean-field and maximum likelihood approach when the sparsity is 0.5, $N = 100$ and $d = 10$. As soon as more observations are available, all methods perform approximately the same. The results are similar in the second row, albeit starker: when the amount of data points are low and sparsity is high, the NF approach can handily outperform the other approaches, see the $N = 100, d = 50$ cases. When multicollinearity is high ($\rho = 0.9$) and selecting incorrect covariates that correlate strongly with the truly relevant ones can lead to similar predictive performance, this advantage becomes smaller. Under lower multicollinearity ($\rho = 0.5$), the NF approach is often far better than competing approaches, as correct selection of covariates is more important.

When the problem becomes more high-dimensional in the third row ($d=200$), the results are similar again, with some caveats. In the lower multicollinearity setting, all methods perform poorly, leading to roughly equal predictive performance. This is the result of all methods predicting the marginal distribution of the data for new data points, as there is too much noise to effectively filter out signal. The only exception lies in the $N = 2500, d = 200$ scenario with sparsity = 0.9, where the aggressive shrinkage of the triple gamma with $a = c = 0.1$ can effectively filter out irrelevant covariates and thereby give better predictions than competing methods. In the higher multicollinearity setting, the NF approach is able to consistently outperform the competing methods, with this advantage becoming larger the sparser the true underlying covariates are. 

The results seem to imply that the combination of strong shrinkage in the kernel as well as the possibility to approximate complex posteriors is most advantageous when the conditions are adverse, namely in situations that are high-dimensional, highly sparse and with covariates that are strongly correlated with one another. As all methods perform similarly to one another when conditions are more favorable, there seems to be little disadvantage in using the proposed approach over others, making it a fairly "safe" choice for estimating GPR models.

\section{Application to BACE-1 Inhibitor Data} \label{sec:app}

\begin{figure}[ht!]
    \centering
    \includegraphics[width=0.95\textwidth]{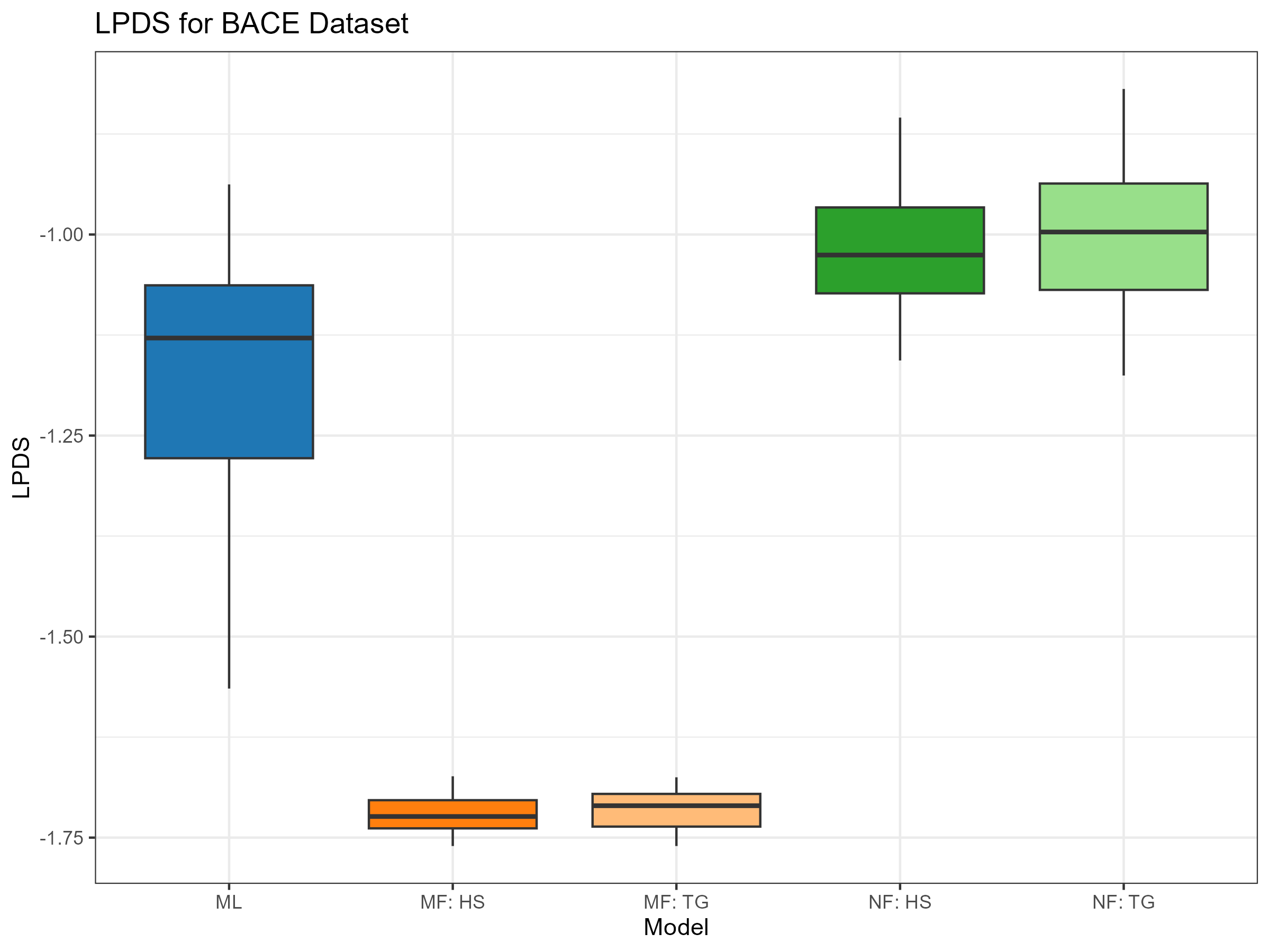}
    \caption{Log Predictive Density Scores (LPDS) across 10-fold cross-validation for Maximum Likelihood (ML), mean-field (MF) approximation under a horseshoe (MF: HS) and triple gamma prior (MF: TG) and the proposed approach under a horseshoe prior (NF: HS) and triple gamma prior (NF: TG) for the BACE-1 inhibitor dataset}
    \label{fig:BACE_results}
\end{figure}

\begin{figure}[ht!]
  \centering
  \includegraphics[width=0.95\textwidth]{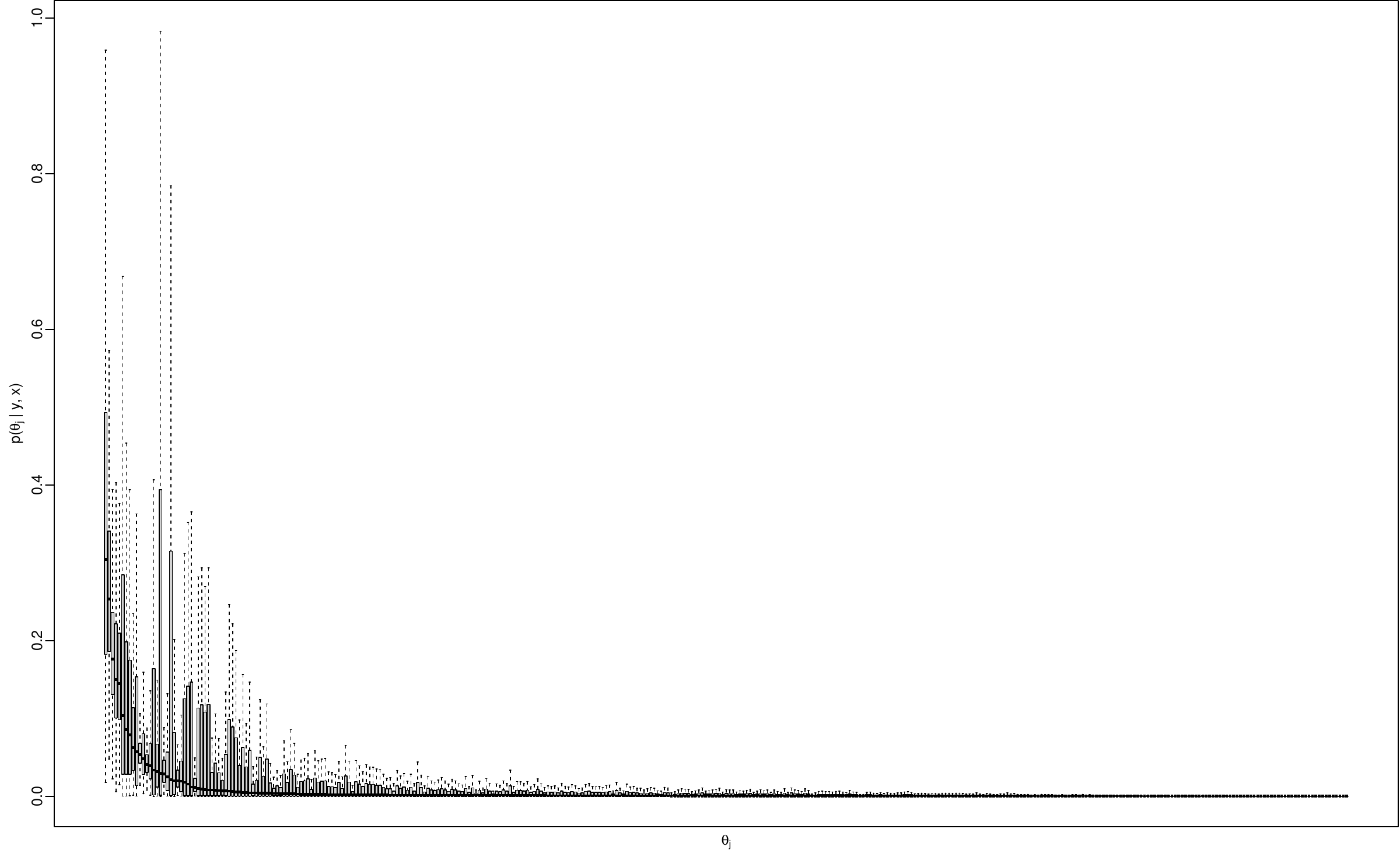}
  \caption{Posterior distribution of the model parameters $\theta_1, \dots, \theta_d$ for the BACE-1 inhibitor dataset under triple gamma prior with $a = c = 0.1$. Sorted in descending order of posterior median.}
  \label{fig:BACE_posterior}
\end{figure}

$\beta$-secretase (BACE-1) is widely recognized as a key therapeutic target for Alzheimer's disease \citep{vassar2011beta}, as the inhibition of BACE-1 has been shown to reduce the formation of amyloid plaques in the brain \citep{rossner2001neuronal}, which are associated with the disease. The BACE-1 inhibitor dataset \citep{wu2018moleculenet} contains  quantitative binding results for 1513 potential inhibitors of BACE-1, described by 589 molecular descriptors. The target is to predict the $\text{pIC}_{50}$ value, which is the negative logarithm of $\text{IC}_{50}$, which in turn is the concentration of an inhibitory substance (e.g., a drug) required to reduce a specific biological process or component's activity by 50\% in vitro.

The dataset was preprocessed by removing all covariates with a variance of zero, scaling all non-binary covariates to have a mean of zero and a standard deviation of one and centering the response variable. This left 363 explanatory variables. All models were evaluated by their out-of-sample LPDS based on 10-fold cross-validation, with each training set containing 90\% of the data and the corresponding test set containing 10\%. The models compared are the same as those used in the simulation study, with the only difference being the amount of iterations allowed for convergence for both variational inference approaches. The normalizing flow models were run for 5000 iterations to ensure convergence, although this was typically achieved earlier. The mean-field approach was run for 10000 iterations, showing worse convergence than the normalizing flow approach.

The results are presented in Figure~\ref{fig:BACE_results}. The proposed normalizing flow approach outperforms both the maximum likelihood and mean-field method, especially when using the more aggressive triple gamma prior. The mean-field approach often did not converge to a useful result, running the maximum number of iterations with the ELBO stagnating or even diverging. To gain some insight into the results generated by the normalizing flow approach, the posterior distribution of $\theta_1, \dots, \theta_d$ under the triple gamma prior with $a = c = 0.1$ are visualized in Figure~\ref{fig:BACE_posterior}, sorted in descending order of posterior median. Many of the $\theta_j$ have a posterior that concentrates around zero, indicating that the corresponding covariates do not meaningfully contribute to the model. This is in line with the expectation that many molecular descriptors in the dataset are not directly related to BACE-1 inhibition. On the flip side, the parameters that are further away from zero often display some uncertainty and, when looking at pair plots, dependencies between them can be observed. Such intricacies could not be accurately captured by either competing method and most likely contributed to the better predictive performance of the proposed approach.

\section{Concluding Remarks and Outlook} \label{sec:conclusion}

This work has contributed to the literature on GPR in two key ways. First, hierarchical shrinkage in the form of the hierarchical triple gamma prior was investigated as a mechanism to mitigate the curse of dimensionality in high-dimensional GPR settings. The prior demonstrated the ability to effectively exclude irrelevant covariates while maintaining flexibility in model size. Second, a novel approach to estimating GPR models was introduced, leveraging normalizing flows within a variational inference framework. This approach enables the approximation of complex posterior distributions in a computationally efficient manner, capitalizing on advances in GPU computing. The proposed methodology outperformed a standard maximum likelihood-based approach as well as a mean-field based approach in simulation studies, particularly in settings characterized by high dimensionality and sparsity. This advantage was also borne out in an application to BACE-1 inhibitor data, which is charaterized by a large amount of covariates and high sparsity. Here, the combination of normalizing flows and aggressive shrinkage was instrumental in providing increased predictive performance. Additionally, the methods presented here are implemented in the R package \texttt{shrinkGPR} \citep{shrinkGPR}, which is publicly available on CRAN.

The use of normalizing flows for posterior approximation remains a relatively novel concept in the context of GPR models, and there are several promising avenues for future research. First, the choice of transformation is critical to the method's performance, and exploring alternatives such as the Inverse Autoregressive Flow \citep{kingma2016improved} could yield further improvements. Recent studies, including \cite{kong2020expressive}, suggest that Sylvester flows may lack the expressivity of other options, underscoring the potential benefits of investigating more flexible transformations. Second, while the methodolgy is currently able to scale to datasets of reasonable size, extending it to sparse GPR models \citep[see, e.g.,][for an overview]{quinonero2005unifying} could enable its application to truly large datasets. However, such an extension would need to address the computational challenges associated with estimating a large number of inducing points, which would significantly increase the dimension of the posterior, thereby also increasing the size of the approximating normalizing flow. Finally, extending the approach to non-Gaussian outcomes may be of interest. The absence of a closed-form marginalized likelihood in this setting would necessitate inference of the latent function $f(\mathbf{x})$, increasing posterior dimensionality and potentially reducing computational efficiency. These challenges notwithstanding, the proposed methodology offers a promising foundation for future research in GPR estimation.

\section{Disclosure statement}\label{disclosure-statement}

The author declares that they have no competing interests.

\section{Data Availability Statement}\label{data-availability-statement}

The BACE-1 data is, as of 2025, publicly available at \url{https://moleculenet.org/datasets-1}.

\phantomsection\label{supplementary-material}
\bigskip

\begin{center}

{\large\bf SUPPLEMENTARY MATERIAL}

\end{center}

\begin{description}
\item[R-package \texttt{shrinkGPR}:]
R-package \texttt{shrinkGPR}, which implements the methods described in the article, is available on CRAN at \url{https://CRAN.R-project.org/package=shrinkGPR}.

\end{description}

\bibliography{refs_jcgs}

\end{document}